\documentclass[aps,prl,showpacs,twocolumn,floatfix,superscriptaddress]{revtex4}
\usepackage{graphicx}
\usepackage{amsmath}
\usepackage{epstopdf}
\usepackage{mathbbol}
\DeclareMathAlphabet\mathbfcal{OMS}{cmsy}{b}{n}
\usepackage{color}

\begin{document}


\title{Large-scales patterns in a minimal cognitive flocking model: \\ incidental leaders, nematic patterns, and aggregates}

\date{\today}

\author{Lucas Barberis}
\affiliation{Universit{\'e} Nice Sophia Antipolis, Laboratoire J.A. Dieudonn{\'e}, UMR 7351  CNRS, Parc Valrose, F-06108 Nice Cedex 02, France}
\affiliation{IFEG, FaMAF, CoNICET, UNC, C{\'o}rdoba, Argentina}
\author{Fernando Peruani}
\email{peruani@unice.fr}
\affiliation{Universit{\'e} Nice Sophia Antipolis, Laboratoire J.A. Dieudonn{\'e}, UMR 7351  CNRS, Parc Valrose, F-06108 Nice Cedex 02, France}

\begin{abstract}
We study a minimal cognitive flocking model, which assumes that the moving entities navigate using exclusively the available instantaneous visual information. The model consists of  active particles, with no memory, that interact by a short-ranged, position-based, attractive force that acts inside a {\it vision cone} (VC) and lack velocity-velocity alignment. 
We show that this active system 
 can exhibit -- due to the VC that breaks Newton's third law -- various complex, large-scale, self-organized patterns.  
Depending on parameter values, we observe the emergence of  aggregates or milling-like patterns, the formation of moving 
 -- locally polar -- files with particles at the front of these structures acting as effective leaders, and the self-organization of particles into  macroscopic nematic structures leading to long-ranged nematic order.
Combining simulations and non-linear field equations, we show that position-based active models, as the one analyzed here, represent a new class of active systems    
fundamentally different from other active systems, including velocity-alignment-based flocking systems. 
The reported results are of prime importance in the study, interpretation, and modeling of collective motion patterns in  living and non-living active systems.
\end{abstract}



\maketitle


It is believed that complex, self-organized,  collective motion patterns observed 
in  birds, fish, or sheep~\cite{vicsek2012,marchetti2013,ballerini2008,gautrais2012,ginelli2015,toulet2015} as well as non-living active systems~\cite{grossman2008,deseigne2010,weber2013,dauchot2015}
 result from the presence of a velocity alignment mechanism that mediates  the interactions among the moving individuals.    
Such velocity alignment mechanism is at the core of the so-called Vicsek-like models~\cite{vicsek1995}  
extensively used to study flocking patterns~\cite{vicsek2012,marchetti2013}. 
Intrinsically non-equilibrium, these patterns differ remarkably from those observed in equilibrium systems 
by the lack of both, Galilean invariance and momentum conservation, 
which allows, for instance, the emergence of long-range orientational order in two-dimensions~\cite{vicsek1995,toner1995,toner1998} and 
the presence of anomalous  density fluctuations~\cite{ramaswamy2003,ramaswamy2010}.

\begin{figure}
\begin{center}
\resizebox{\columnwidth}{!} {\includegraphics{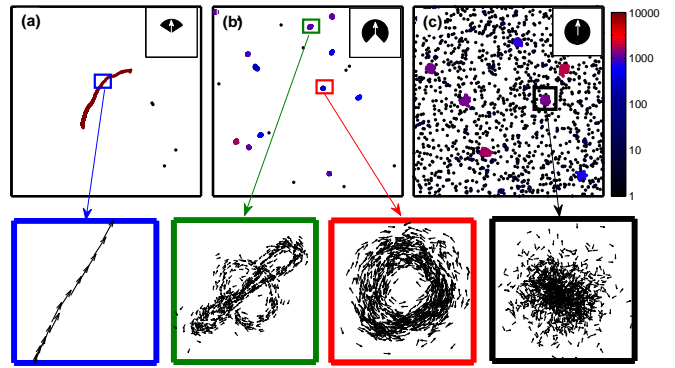}}
\caption{
(Color online) Depending on vision cone size $\beta$ (insets), we can observe 
the formation of locally polar structures, which we call ``worms", panel (a) -- $\beta=0.8$, 
or the emergence of  aggregates/milling-like patterns as shown in panel (b) for $\beta=2.4$, 
and panel (c) for  $\beta \sim \pi$. Parameters (a)-(c): $\sqrt{2 D_{\theta}}=0.12$, $L=100$, $N=10^4$.
See~\cite{NoteSI} for movies. 
 }
\label{fig:AggWorms}
\end{center}
\end{figure}

Few recent pioneering works~\cite{romanczuk2009,strombom2011,moussaid2011,pearce2014,huepe2013,huepe2015,grossmann2013,soto2014} have challenged the wide-spread view that behind each collective motion pattern of self-propelled entities, there is a velocity alignment mechanism at work. 
Here, we explore the possibility of observing flocking patterns in the absence of such alignment.  
%
The model we analyze is a minimal cognitive flocking model that assumes that the moving entities navigate 
 using exclusively the instantaneous visual information they receive.  
Importantly, the moving particles have no memory as to compute the moving direction of neighboring particles, in sharp contrast to standard flocking models~\cite{vicsek1995,vicsek2012,marchetti2013}. 
The navigation strategy we investigate is based on the instantaneous position of neighboring particles and not on their velocity, 
which makes the model simpler from a cognitive point of view and computationally less intensive, providing an alternative in the design of robotic navigation algorithms.
The model incorporates few well-known physiological and cognitive concepts. 
For instance, we assume that particles are attracted by those particles located inside the {\it vision cone}. 
The {\it vision cone} (VC) results from two well-documented facts:    
i) animals have a limited field of view~\cite{moussaid2011,gibson1958,mccomb2008} --  typically less than 360 degree --, which is parametrized here by the angle $\beta$,  
and ii) when navigating, objects located at far distances are ignored and the focus is put on those objects located at distances shorter 
than the so-called {\it cognitive horizon}~\cite{moussaid2011,gibson1958,strandburg2013} that corresponds 
in our model to $R_0$.
The field of view ($\beta$) is known 
to vary from species to species, being for instance smaller for predators than for  preys~\cite{gibson1958,mccomb2008}.  
In summary, the field of view ($\beta$) combined with the cognitive horizon ($R_0)$ define the VC, 
which violates Newton's third law for $\beta<\pi$. 
Notice that there exist alternative mechanisms to break the action-reaction symmetry to the VC~\cite{soto2014,dzubiella2003,ivlev2015}  
 and that the presence of non-reciprocal interactions has also a strong impact on the dynamics of flocking models with velocity alignment~\cite{dadhichi2016,cavagna2016,nguyen2015,durven2016}.

Here, we show that this minimal cognitive flocking model exhibits various large-scale self-organized patterns, depending on the size of the VC and noise intensity: 
aggregates or milling-like patterns of various degrees of complexity,  locally polar dynamical structures which we call worms (Fig.~\ref{fig:AggWorms}), 
and nematic bands leading to long-ranged nematic order. 
Furthermore, we derive a system of non-linear field equations to rationalize agent-based simulation results and show 
 that in general position-based active models, as the one here studied, represent a new class of active systems    
fundamentally different from other active systems, including velocity-alignment-based flocking systems~\cite{vicsek1995,toner1995,toner1998,chate2008,toner2012,ramaswamy2003,chate2006,ngo2014,peruani2006,peruani2008,baskaran2008,ginelli2010,peshkov2012b,abkenar2013,weitz2015,nishiguchi2016}.

\begin{figure}
\begin{center}
\resizebox{\columnwidth}{!} {\includegraphics{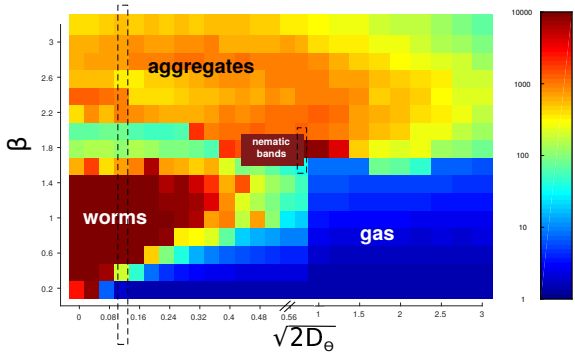}}
\caption{
(Color online)  Phase diagram: vision angle $\beta$ vs angular noise intensity $D_{\theta}$. The color code indicates  
the value of the average cluster size $m^*$ in simulations with $N=10^4$. Vertical rectangles refer to cuts of the phase diagram shown 
in Fig.~\ref{fig:worm} and Fig.~\ref{fig:nematic}.
 }
\label{fig:PD}
\end{center}
\end{figure}

{\it Model definition.--}
The equation of motion of the $i$-th particle is given by:
\begin{eqnarray}
\label{eq:posvel}
\begin{array} {cc}
\dot{\mathbf{x}}_i  =  v_0 \mathbf{V}(\theta_i) ; & 
  \dot{\theta}_i =  \frac{\gamma}{n_i} \sum_{j \in \Omega_i} \sin(\alpha_{ij}-\theta_i)+ \sqrt{2D_{\theta}}\xi_i(t) \label{eq:vel}\! \, ,
\end{array}
\end{eqnarray}
where $\mathbf{x}_i$ denotes the position of the particle, $\theta_i$ represents its moving direction, with  $\mathbf{V}(.)\equiv (\cos(.),\sin(.))^{T}$, 
$v_0$ is the particle speed, $\gamma$ the strength of the interactions, and $\xi_i(t)$ is a noise term such that $\langle \xi_i(t) \rangle = 0$ and  $\langle \xi_i(t) \xi_j(t') \rangle = \delta_{i,j} \delta(t-t')$, with the noise amplitude given by $D_{\theta}$. 
The sum  in Eq.~(\ref{eq:vel})  describes the projection on the ``retina'' of particle $i$ of the position of all particles inside its VC, assuming particles are point-like, with $\alpha_{ij}$ the polar angle of the vector $\frac{\mathbf{x}_j - \mathbf{x}_i}{||\mathbf{x}_j - \mathbf{x}_i||}=\mathbf{V}(\alpha_{ij})$; 
a procedure  similar to the one in~\cite{pearce2014} for long-range interactions.  
The symbol $\Omega_i$ thus denotes the set of neighbors inside the VC of  particle $i$, with $n_i$ its  
cardinal number.  
Particles in  $\Omega_i$ are those that satisfy  $||\mathbf{x}_j - \mathbf{x}_i||\leq R_0$ 
and  $\frac{\mathbf{x}_j - \mathbf{x}_i}{||\mathbf{x}_j - \mathbf{x}_i||}\cdot\left({\dot{\mathbf{x}}_i}/{||\dot{\mathbf{x}}_i||}\right)>\cos(\beta)$, with $\beta$ the {\it size} of the cone
and its orientation given by $\dot{\mathbf{x}}_i$. 
%
%
For a definition of the model in 3D and a justification of the term $\sin(\alpha_{ij}-\theta_i)$, see~\cite{NoteSI}. 
In the following we fix $v_0=1$, set  $R_0=1$, $\gamma=5$, and the global density $\rho_0=N/L^2=1$, with $N$ the number of particles and $L$ the linear size of the system 
and use periodic boundary conditions. These parameters are in the range of the ones expected for vertebrates~\cite{noteBioParam}.  

\begin{figure}
\begin{center}
\resizebox{\columnwidth}{!} {\includegraphics{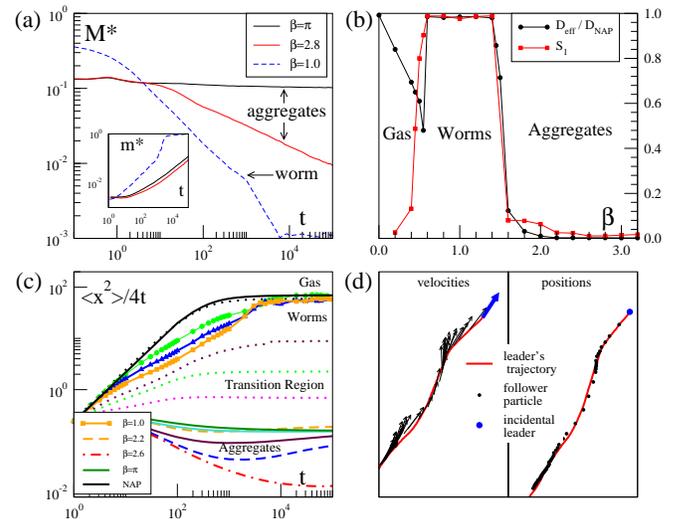}}
\caption{
(Color online)  
(a) Normalized number of clusters $M^*$ (normalized average cluster size $m^*$, inset) as function of time. 
Worms coagulate at a much faster rate 
than aggregates. 
%
(b) The (normalized) diffusion coefficient $D_{eff}/D_{NAP}$ and polar order $S_1$ as function of $\beta$ (see text). 
(c) Temporal evolution of $\langle \mathbf{x}^2\rangle /(4t)$. 
(d) Velocities and positions of particles that form a worm:  
  particles  
copy the behavior of the particle at the front, which we refer to as the {\it incidental leader}.
%
See~\cite{NoteSI} for a movie. 
Parameters: $N=10^4$, $L=100$, $\sqrt{2 D_{\theta}}=0.12$.
 }
\label{fig:worm}
\end{center}
\end{figure}

\begin{figure}
\begin{center}
\resizebox{\columnwidth}{!} {\includegraphics{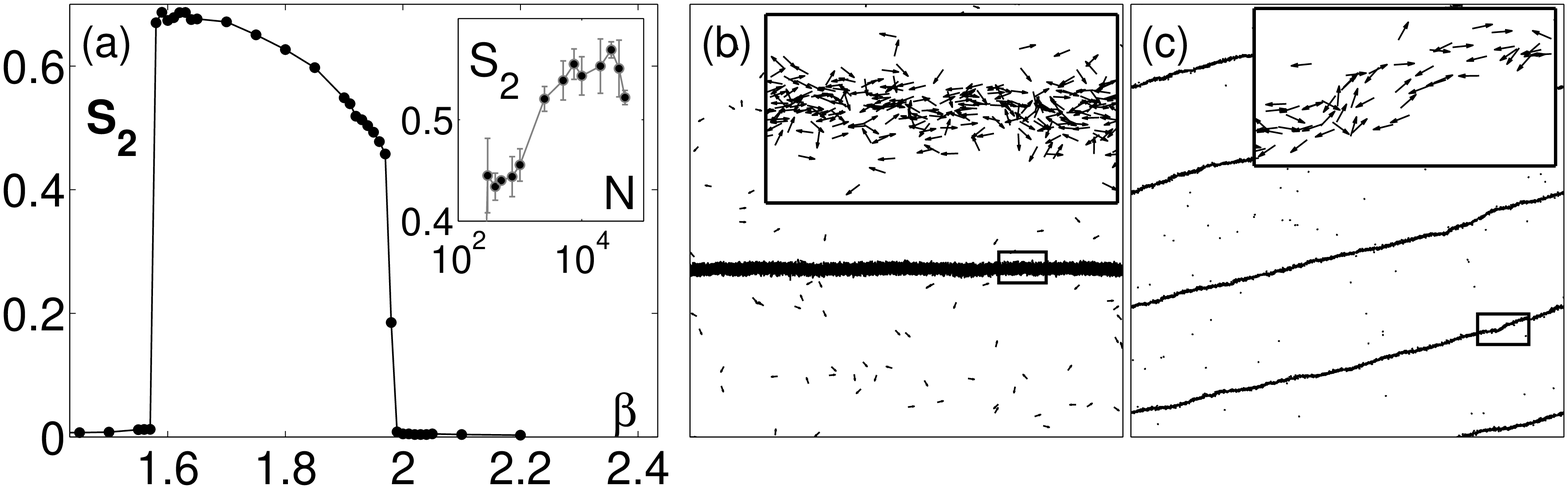}}
\caption{
Nematic bands. (a) The nematic order parameter $S_2$ as function of $\beta$ for $\sqrt{2 D_{\theta}}=0.84$. The inset shows  $S_2$ as function of $N$, 
with error bars obtained using $50$ realizations, $\beta=1.9$. 
(b) and (c) display simulations snapshots at the steady state for $N=10^4$, $L=100$, $\sqrt{2 D_{\theta}}=0.84$, and $\beta=1.9$. Insets in (b) and (c) correspond to magnified views of the bands. See~\cite{NoteSI} for movies.  
 }
\label{fig:nematic}
\end{center}
\end{figure}

{\it Phenomenology.--}
The system exhibits four distinct phases -- see Fig.~\ref{fig:PD} -- which we refer to as 
i)  gas phase, ii)  aggregate phase, iii) worm phase, and iv) nematic phase. 
Phases ii) to iv) involve spontaneous phase separation of the particles, 
while phase i) is characterized by the absence of order and an homogeneous distribution of particles in space.    
In the following, we study the phase-separated phases, i.e. from ii) to iv), by performing two vertical cuts in the phase diagram in Fig.~\ref{fig:PD}. 
%
%
The emerging macroscopic patterns are characterized by their level of (global) orientational order 
through $S_q   =   | \sum_{j=1}^{N} \exp(\imath \,q\theta_j)/N |$, 
%
with $q=1$ for polar order and $q=2$ for nematic order (and $\imath$ the imaginary unit). 
 Clustering properties are analyzed by the normalized number of clusters $M^* = \langle M \rangle/N$ and normalize cluster size $m^* = \langle m \rangle/N$, defining a cluster a set of connected particles, where  $i$ is connected to  $j$ if $j$ is located inside the VC of $i$. 
Finally,  transport properties are studied by looking at the behavior of the  diffusion coefficient $D_{eff} = \lim_{t\to\infty} \sum_{i=1}^N \left(\mathbf{x}_i(t) - \mathbf{x}_i(t_0)\right)^2/[4 N (t-t_0)] $. 

{\it Aggregate phase.--} 
At large values of $\beta$, particles self-organize into aggregates of different complexity,  Fig.~\ref{fig:AggWorms}, 
 with some of these patterns comparable to the ones reported in~\cite{dorsogna2006}.  
The aggregates result from a phase separation process (see~\cite{NoteSI} for movies) fundamentally different from the one in~\cite{tailleur2008,marchetti2012b}. 
%
Fig.~\ref{fig:worm}(a) shows the existence of different nontrivial scalings of $M^*$ and $m^*$ with time -- as expected for active systems~\cite{mones2015} -- which suggests that the phase-separation process is of a different nature at large and intermediate values of $\beta$.

{\it Worm phase.--} 
At  lower values of $\beta$, we observe the emergence of a new type of macroscopic structure, which we call worm,  Fig.~\ref{fig:AggWorms}(a). 
%
This structure consists of a file of active particles that are locally polarly oriented.
%
The particle at  the ``head" of the worm -- which we label ``$H$'' -- ignores all other particles and becomes the effective leader of the spontaneously formed herd of active particles. 
This is evident from the behavior of $D_{eff}$ as shown in Fig.~\ref{fig:worm}. 
During the worm phase, $D_{eff} \sim  D_{NAP} = v_0^2/(2D_{\theta})$, 
with $D_{NAP}$ the diffusion coefficient of an ensemble of non-interacting active particles.
This is not due to the absence of interaction, but to the fact that all particles in the worm imitate the behavior of the incidental leader particle: the position and velocity of  particle $j$  is approximately given by $\mathbf{x}_j(t) \sim \mathbf{x}_H(t-\ell_{j,H}/v_0)$ and $\mathbf{V}(\theta_j(t)) \sim \mathbf{V}(\theta_{H}(t-\ell_{j,H}/v_0))$, where $\ell_{j,H}$ is the distance along the worm between $j$ and $H$, Fig.~\ref{fig:worm}(d) and~\cite{NoteSI} for a movie.
Though worms exhibit local polar order,  global polar order drops for $L \gg v_0/D_{\theta}$, vanishing in the thermodynamic limit.  

{\it Nematic phase.--} 
For larger values of $D_{\theta}$  and $\beta$, see Figs.~\ref{fig:PD} and~\ref{fig:nematic}, we find  macroscopic nematic bands. 
After a complex transient where various small nematic bands grow in size and interconnect, the system reaches a steady state, with one or several bands, but where only one direction prevails, 
see Fig.~\ref{fig:nematic} and~\cite{NoteSI} for a movie. 
The described dynamics leads to the emergence of genuine global nematic order.  Increasing the system size $N$, for a fixed density $\rho_0$, we observe that  the nematic order $S_2$ saturates,  Fig.~\ref{fig:nematic}(a), inset.

{\it Field equations.--} 
A qualitative understanding of the  large-scale behavior of the system can be obtained 
in terms of $p(\mathbf{x}, \theta, t) = \langle \sum_{i=1}^N \delta(\mathbf{x} - \mathbf{x}_i) \delta(\theta - \theta_i) \rangle$. 
The evolution of $p(\mathbf{x}, \theta, t)$ is given by the corresponding non-linear Fokker-Planck equation of Eq.~(\ref{eq:posvel})~\cite{risken}:
\begin{eqnarray}
\label{eq:FP}
 \partial_t p + \mathbf{\nabla} \left[ v_0 \mathbf{V}(\theta) p\right]   =  D_{\theta} \partial_{\theta \theta} p - \partial_{\theta}\left[ \mathcal{I} p \right]\, ,
\end{eqnarray}
where we have assumed that  $p_2(\mathbf{x}, \theta, \mathbf{x}', \theta', t) \simeq p(\mathbf{x}, \theta,t) p(\mathbf{x}', \theta',t)$ in 
order to define, after some simple calculations~\cite{NoteSI}, an average interaction term: 
\begin{eqnarray}
\label{eq:Interaction}
\mathcal{I} =  \Gamma \int_{0}^{R_0} dR  \int_{\theta-\beta}^{\theta+\beta} d\alpha R\, \sin(\alpha - \theta) \rho(\mathbf{x} + R \mathbf{V}(\alpha),t) \,,
\end{eqnarray} 
where $\Gamma \simeq \gamma/(1+\beta  R_0^2 \rho)$ and $\rho(\mathbf{x} ,t) = \mathcal{L}[1]$ the coarse-grained one-particle density, with $\mathcal{L}[(.)]$ an averaging operator defined as $\mathcal{L}[(.)] \equiv \int_0^{2\pi} d\theta \, (.) p(\mathbf{x}, \theta, t)$. 
To analyze the behavior of Eq.~(\ref{eq:FP}), in addition to $\rho(\mathbf{x} ,t)$ we introduce the following fields: local polar order $\mathbf{P} \equiv (P_x,P_y)^T = \mathcal{L}[\mathbf{V}(\theta)]$,  local nematic order 
$\mathbf{Q} \equiv (Q_c, Q_s)^T = \mathcal{L}[\mathbf{V}(2\theta)]$, and higher order fields are denoted by $\mathbf{M}_k \equiv (M_{kc},M_{ks})^T = \mathcal{L}[\mathbf{V}(k\theta)]$ where $k>2$ corresponds to the local $k$-th order field, and recast  the equation as: 
\begin{widetext}
\begin{subequations}
\label{eq:fields}
\begin{align}
\label{eq:fields_1}
\partial_t \rho+v_0 \nabla \mathbf{P}&=0 \\
\label{eq:fields_2}
\partial_t \mathbf{P} + \frac{v_0}{2} \left(\nabla \rho + \left[ \nabla^T \overline{\overline{\mathcal{M}}}_Q\right]^T \right) &= -D_{\theta} \mathbf{P}  
 - \frac{\Gamma g(\beta)}{2} \left[  \overline{\overline{\mathcal{M}}}_Q - \rho \mathbb{1}  \right] \nabla \rho 
- \frac{\Gamma f(\beta)}{2} \overline{\overline{\mathcal{M}}}_{\rho1} \left[ \mathbf{P} - \mathbf{M}_3 \right] \\
 \label{eq:fields_3}
 \partial_t \mathbf{Q} + \frac{v_0}{2} \left[\mathbf{\nabla}^T \left(
\overline{\overline{\mathcal{M}}}_3 + \overline{\overline{\mathcal{M}}}_P  
 \right)\right]^T &=
 -4 D_{\theta} \mathbf{Q} 
 - \Gamma g(\beta) \left[ \overline{\overline{\mathcal{M}}}_3 - \overline{\overline{\mathcal{M}}}_P^{T} \right]  \nabla \rho 
 - \Gamma f(\beta) \left( \overline{\overline{\mathcal{M}}}_{\rho2} \mathbf{M}_4 +  \rho 
\left[ 
\begin{array}{c} 
\Phi \rho \\ 
-\partial_{xy}\rho 
 \end{array}
\right] \right) \, ,
  \end{align} 
\end{subequations}
\end{widetext}
where the symbols $\overline{\overline{\mathcal{M}}}_A$ denote matrices defined 
using the auxiliary matrices  $\mathbb{E}_1 = \left[ \begin{array}{cc} 1 & 0 \\ 0 & -1 \end{array} \right]$, 
$\mathbb{E}_2 = \left[ \begin{array}{cc} 0 & 1 \\ 1 & 0 \end{array}  \right]$, $\mathbb{E}_3 = \left[ \begin{array}{cc} 0 & 1 \\ -1 & 0  \end{array} \right]$, and the unity matrix $\mathbb{1}$ as: 
$\overline{\overline{\mathcal{M}}}_Q = Q_c \mathbb{E}_1 + Q_s \mathbb{E}_2$, $\overline{\overline{\mathcal{M}}}_3 = M_{3c} \mathbb{E}_1 + M_{3s} \mathbb{E}_2$, 
$\overline{\overline{\mathcal{M}}}_P = P_x  \mathbb{E}_2 +  P_y  \mathbb{E}_3$, $\overline{\overline{\mathcal{M}}}_{\rho1} = \Phi \rho/2  \mathbb{E}_1 - \partial_{xy}\rho   \mathbb{1}$, and 
$\overline{\overline{\mathcal{M}}}_{\rho2} = \partial_{xy} \rho \mathbb{E}_2 - \Phi\rho/2   \mathbb{E}_1$. 
In addition, we have defined $\Phi \rho$ as $\Phi \rho = \partial_{yy} \rho - \partial_{xx} \rho$ and  
the terms $g(\beta)$ and $f(\beta)$ are, respectively, the first and second non-zero terms in the expansion of $\mathcal{I}$ with respect $R_0$, that read  
$g(\beta) = (R_0^3/3) \left( \beta - \sin(2 \beta)/2 \right)$ and $f(\beta) = (R_0^4/6) \sin^3(\beta)$. 
Equations~(\ref{eq:fields}), due to the $k>2$ order fields, require a closure ansatz. 
This can be done by providing an ansatz on the local order as explained below.


\noindent For $\beta \sim \pi$, i.e. (quasi) isotropic  interactions, $f \simeq0$, and assume that no local order is possible. 
Under these conditions, Eqs.~(\ref{eq:fields}) reduce to:
\begin{eqnarray}
\label{eq:phaseSep}
\partial_t \rho =  - \frac{v_0}{D_{\theta}} \nabla\left[ - \frac{v_0}{2} \nabla \rho + \Gamma g(\beta) \rho \nabla \rho + \mathcal{O}(R_0^5) \right] \, ,
\end{eqnarray} 
where $\mathcal{O}(R_0^5)$ contains spatial third order derivatives with respect to $\rho$. 
From Eq.~(\ref{eq:phaseSep}) we learn that an homogenous spatial distribution of particles --  assume $\rho = \rho_0 + \epsilon \delta \rho$, with $\rho_0$ a constant and $\epsilon \delta \rho$ a small perturbation -- becomes  linearly unstable when $c_1 =   v_0 \Gamma g(\beta)\rho_0/(2 D_{\theta}) - D_{NAP} >0 $  and  the system undergoes  phase separation, in absence of orientational order, that  
leads to the emergence of aggregates as shown in Fig.~\ref{fig:AggWorms}, panels (b) and (c). For $\beta=\pi$, the dispersion relation is of the form $\lambda =  c_1 \mathbf{k}^2 - c_2 \mathbf{k}^4$, where $c_2 = v_0 \Gamma \pi R_0^5 \rho_0/80 >0$ and  $\mathbf{k}$ the wavenumber associate to the perturbation.

For intermediate values of $\beta$ and $D_{\theta}$  agent-based simulations display nematic patterns. 
Let us then  assume that locally the distribution of $\theta$ is given by  
$p(\mathbf{x}, \theta, t) \simeq (\rho/2\pi) \exp[(2/\rho)\mathbf{Q}.\mathbf{V}(2\theta)]$~(see~\cite{NoteSI} for a derivation). 
As direct consequence of this local ansatz we find that $\mathbf{P}=\mathbf{M}_3=\mathbf{0}$, $M_{4c} = \left( Q_c^2 - Q_s^2 \right)/(2\rho)$ and $M_{4c} = Q_c Q_s/\rho$. Since under this assumption $\mathbf{M}_{4}$ can be expressed in terms of $\rho$ and $\mathbf{Q}$, Eqs.~(\ref{eq:fields}) define a closed set of equations. Now we look for the stationary states of the resulting system. This implies that all partial temporal derivatives of the fields vanish. For simplicity, but without loss of generality, let us assume that $Q_s=0$ and that the system is invariant in the $\hat{x}$ direction as in Fig.~\ref{fig:nematic}.c, and thus derivatives in $x$ vanish. 
Inserting all this into  Eqs.~(\ref{eq:fields}), we arrive to: 
\begin{subequations}
\label{eq:band}
\begin{align}
\label{eq:band:1}
\partial_y \left( \rho - Q_c \right) &=  - \frac{\Gamma g(\beta)}{v_0} \partial_y \rho \left(\rho + Q_c \right) \\
\label{eq:band:2}
Q_c &= - \frac{\Gamma f(\beta)}{8 D_{\theta}} \partial_{yy} \rho \left( \rho - \frac{Q_c^2}{2\rho} \right) \, ,
\end{align}
\end{subequations} 
where $Q_c$ and $\rho$ are functions of $y$. 
By linearizing this system of equations -- assume $\rho = \rho_0 + \epsilon \delta \rho$ and $Q_c = \epsilon \delta Q_c$, with $\rho_0$ a constant and  $\epsilon \delta \rho$ and  $\epsilon \delta Q_c$ perturbations in the density and nematic order, respectively, and keeping linear order terms in $\epsilon$ --  it becomes evident that $\delta Q_c \propto \partial_{yy}\delta\rho$ and the system reduces to  
$(a \rho_0 - 1)/(b \rho_0) z = \partial_{yy} z$, with $z=\partial_y \delta \rho$, $a=\Gamma g/v_0$ and $b=\Gamma f/(8 D_{\theta})$, 
whose solutions for $ \rho_0 - 1<0$ correspond to trigonometric functions. 
All this means that by assuming local nematic order, we can show that: i) Eqs.~ (\ref{eq:fields}) exhibit  steady state solutions, and 
ii) that these static solutions correspond to elongated high density regions, nematically ordered, with $Q_c \propto \rho$, parallel to each other and equally spaced, {\it i.e.} there is a well-defined wave length. These solutions are consistent with the 
nematic bands in Fig.~\ref{fig:nematic}. 

Finally, we have observed in agent-based simulations locally polar patterns (worms). 
Let us assume then that  
 $p(\mathbf{x}, \theta, t) \simeq (\rho/2\pi) \exp[(2/\rho)\mathbf{P}.\mathbf{V}(\theta)]$~\cite{NoteSI}. 
Under this assumption, it is possible to show that static polar bands -- {\it i.e.} static straight worms -- cannot exist. The field equations suggest that polar structures never reach a steady state  as observed in simulations, see Fig.~\ref{fig:AggWorms}(a) and~\ref{fig:worm}(c).

{\it Concluding remarks.--} 
%
%
The derived field equations, combined with the presented numerical study, show that 
there exist fundamental differences between velocity alignment-based models, including 
polar fluids~\cite{toner1995,toner1998,chate2008,bertin2009}, active nematics~\cite{ramaswamy2003,chate2006,ngo2014}, and self-propelled rods~\cite{peruani2006,peruani2008,baskaran2008,ginelli2010,peshkov2012,abkenar2013,weitz2015,nishiguchi2016}, on the one hand,  
and position-based models as the one analyzed here on the other hand. 
An evident and fundamental difference, revealed by Eqs.~(\ref{eq:fields}) and confirmed in simulations, 
is that position-based models cannot develop either polar or nematic ordered phases that are spatially homogeneous -- cf. with the well reported spatially-homogeneous ordered phases in (velocity-alignment) flocking models, such as the celebrated Toner-Tu polar phase~\cite{toner1995,toner1998,chate2008,bertin2009} and homogeneous nematic phase~\cite{ramaswamy2003,peruani2008,baskaran2008,ginelli2010,peshkov2012}. In contrast, in position-based active models (orientational) order emerges always, even at short-scales, associated to density instabilities.  
In addition,  worms display local polar order that is parallel (locally) to the band, with a highly dynamical center band line that prevents polar long-range order (LRO) to emerge,  in striking difference with polar bands in Vicsek models, where polar order is orthogonal to the band and long-ranged~\cite{gregoire2004,chate2008,bertin2009}. 
On the other hand, it has been argued that nematic bands, reported in self-propelled rods~\cite{ginelli2010} are unstable~\cite{peshkov2012}. 
This is again in sharp contrast to the nematic bands reported here that remain stable in the thermodynamical limit.
%
%
In summary, position-based active systems as the one presented here belong to a new {\it universality} active class fundamentally different  
to any previously reported  
 active matter classes~\cite{vicsek1995,toner1995,toner1998,chate2008,bertin2009,toner2012,ramaswamy2003,chate2006,ngo2014,peruani2006,peruani2008,baskaran2008,ginelli2010,peshkov2012b,abkenar2013,weitz2015,nishiguchi2016}.

Our results could be relevant to study and interpret collective patterns in animal groups. 
They indicate that several flocking patterns observed in nature, such as the milling-like patterns found in fish~\cite{gautrais2012}, the file formation  reported in sheep herds~\cite{toulet2015}, and the emergence of nematic bands in human crowds and ants~\cite{moussaid2011} could result from 
simple navigation strategies that do not require memory, use exclusively the instantaneous position of neighboring particle, 
and limit the interaction neighborhood by a vision cone. 
These concepts, that lead to navigation strategies that are computationally less intensive than those based on velocity alignment,  could help in the design of new robotic navigation algorithms, as for instance for  phototactic robots~\cite{volpe2015}. 
Extensions of this minimal cognitive model could find applications in other active systems such as chemophoretic  particles~\cite{soto2014,ivlev2015} and chemotactic colloids~\cite{pohl2014,saha2014,liebchen2015,zoettl2016} and organisms~\cite{soon2007,beta2012,beta2013}, among other examples where aggregations patterns have been reported. 
This could require either taking $R_0 \to \infty$ or replacing the interaction cut-off by a slowly decaying function of the distance, 
as well as specializing the model for $\beta=\pi$, which corresponds to the limit of isotropic interactions, though asymmetric interactions 
may also be realistic~\cite{soto2014,ivlev2015,soon2007}. 
%

\begin{acknowledgments}

\paragraph{Acknowledgement}
We thank C. Beta, C. Condat, M. Polin, R. Soto, and G. Volpe for insightful comments that helped us to reshape the text.  
LB ~acknowledges financial support from CONICET and FP from Agence Nationale de la Recherche via Grant ANR-15-CE30-0002-01. 
Simulations were carried out in  CRIMSON and CICADA clusters belonging to OCA  and UNSA, respectively.

\end{acknowledgments}

\bibliographystyle{apsrev}

\end{document}